\documentclass[11pt]{article}
\usepackage{graphicx}
\begin{document}
\begin{center}

{\Large Nanomachines Based on Carbon Nanotubes}
\\

Yu.E. Lozovik\footnote{E-mail: lozovik@isan.troitsk.ru}, A.V. Minogin, A.M.
Popov\footnote{E-mail: popov@ttk.ru}

Institute of Spectroscopy, Russian Academy of Science, 142190, Troitsk, Moscow region,
Russia
\end{center}

\begin{abstract}
Possibility for double-wall carbon nanotube to operate as the bolt-and-nut pair is studied. The
barriers to relative motions of walls along the helical "thread" line and to jumps on neighbouring
helical lines are calculated as functions of the wall lengths for the set of double-wall carbon
nanotubes. The dynamics of the relative motion of carbon nanotube walls along the helical line under
the action of external forces is considered. Perforated nanodrill, variable nanoresistor and other
principally new nanotube based mechanical nanodevices using this motion are proposed. Possible
operation modes of proposed nanodevices are discussed.
\end{abstract}



\section{Introduction}

Progress in nanotechnology in last decades have given rise to the possibility to manipulate with
nanometer-size objects \cite{1}. The principal schemes of nanometer-size machines (nanomachines)
where the controlled motion can be realized are considered \cite{2}. Thus, the search of nanoobjects
that can be used as movable elements of nanomachines is a very actual challenge for development of
nanomechanics. The low frictional relative motion of carbon nanotube walls \cite{3,4,tele} and
unique elastic properties \cite{6} of these walls allows to be considered as promising candidates
for such movable elements. A set of nanomachines based on relative sliding of walls along nanotube
axis or their relative rotation is proposed: a constant-force nanospring \cite{tele}, nanobearings
\cite{bearings} and nanogears \cite{gear} driven by laser electric fields; a mechanical nanoswitch
\cite{switch} operated by electrostatic forces; a gigahertz oscillator \cite{giga}.

All these nanomachines correspond to the case where the corrugation of the interwall interaction
energy has little or no effect on relative motion of nanotube walls. However, all of carbon nanotube
walls have the helical symmetry and this gives the possibility for neighbouring walls of nanotube to
be bolt-and-nut pair. The present work is devoted to principally new type of nanomachines where the
relative motion of nanotube walls occurs along helical "thread" lines. The possibility of controling
this motion by the potential relief of the interwall interaction energy is considered. A theory for
dynamics of relative motion of nanotube walls is developed. Possible types of these nanomachines are
discussed. Two operation modes for these nanomachines are analyzed: Fokker-Planck operation mode,
where a relative motion of walls occurs as diffusion with drift under the action of external forces
and the accelerating operation mode, where the relative motion of walls is controlled by external
forces. The values of the controlling forces corresponding to these modes are estimated.

\section{Potential relief of the interwall interaction energy.}

By potential relief we mean the dependence of the interwall interaction energy $U$ of two
neighbouring nanotube walls on the coordinates describing the relative position of the wall. Such
coordinates are the angle $\phi$ of relative rotation of the wall about the nanotube axis and the
length $z$ of relative displacement of the wall along it. It is convenient to visualize the
potential relief as a map plotted on a cylindrical surface. In the general case a nanotube wall has
a helical symmetry (see, for example, \cite{symmetry}). Therefore, the double-wall nanotubes with
the potential relief having valleys directed along the helical line analogeously to a thread of bolt
were found \cite{relief}. Here we study the properties of nanotubes having the potential relief of
the interwall interaction energy that allows them to be a bolt-and-nut pair. We have restricted
ourselves to the case of double-wall carbon nanotubes. Such nanotubes were produced as by the
standard method of synthesis in arc discharge \cite{iijima1} (see also review \cite{ufn} devoted to
the problems of growth of different carbon nanostructures) as recently from the pea-pod nanotubes by
electron irradiation \cite{tcn1} or by heating \cite{tcn2}.

\begin{figure}
\includegraphics[height=8cm]{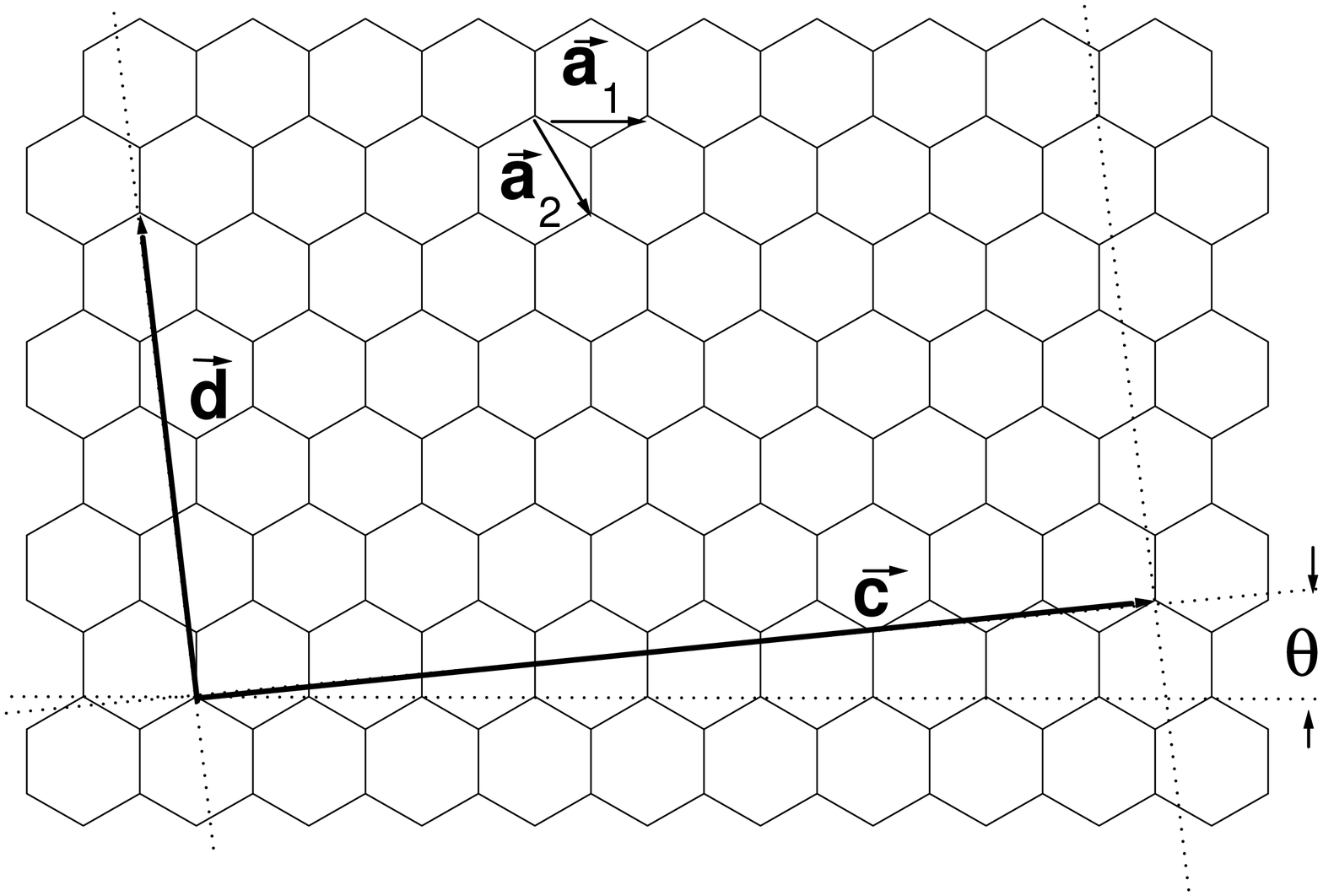}
\caption{\label{fig1_ftt} Graphene sheet used to construct a nanotube wall (see \cite{class}). The
wall is determined by the lattice vector ${\bf c}$. The wall unit vector is denoted by ${\bf d}$.
The chiral angle is denoted by $\theta$. The unit vectors of graphene sheet are denoted by ${\bf
a}_1$ and ${\bf a}_2$.}
\end{figure}

The structure of nanotube wall is determined by the pair of integers $(n,m)$ corresponded to a
lattice vector ${\bf c} = n {\bf a}_1 + m {\bf a}_2$ on the graphene plane used to perform mapping,
where ${\bf a}_1$ and ${\bf a}_2$ are the unit vectors of the graphene sheet \cite{class,class1}
(see Fig 1). The wall radius $R$ is defined as

\begin{equation}
R=\frac{\mid{\bf c}\mid}{2 \pi} = \frac{a_0 \sqrt{(n^2+mn+m^2)}}{2 \pi},
\end{equation}

\noindent where $a_0$ is the length of graphite lattice vector. The length of
the wall unit vector ${\bf d}$ is given by the relation

\begin{equation}
\label{ucl}
      \mid {\bf d} \mid= \frac{3a_0 \sqrt{n^2+mn+m^2}}{GCD(2n+m,2m+n)},
\end{equation}

\noindent where $GCD(q,s)$ is the greatest common divisor of
integers $q$ and $s$. The number of graphite unit cell $N_c$ in a
unit cell of a wall is

\begin{equation}
                  N_c = \frac{2 (n^2+mn+m^2)}{GCD(2n+m,2m+n)},
\end{equation}

The walls of a double-wall nanotube are commensurate if the ratio $d_1/d_2$ of wall unit cell
lengths is a rational fraction and incommensurate otherwise. The chiral angle $\theta$ of wall is
defined as the angle between between vectors ${\bf c}$ and ${\bf a}_1$. This angle equals to

\begin{equation}
\theta = - \arccos \frac{2n+m}{\sqrt{n^2+m^2+mn}}
\end{equation}

Several works were devoted to study of the interwall interaction energy of double-wall carbon
nanotubes [12,19-22]
and double-shell carbon nanoparticles \cite{osawa,my}. The barriers to the relative rotation of the
walls about the nanotube axis and to sliding along it were calculated for the (5,5)@(10,10)
\cite{relief,bar1,bar2}, (7,7)@(12,12) \cite{yug2} and (12,12)@(17,17) \cite{yug2} nanotubes. The
dependencies of barriers to the relative sliding along the nanotube axis on number of atoms in the
nanotube are studied for several commensurate and imcommensurate double-wall nanotubes \cite{along}.
The maps of potential relief of the interwall interaction energy as function of the relative
displacement of the wall along the nanotube axis and angle of the relative rotation of the wall were
considered by Dresselhaus {\it et. al.} for a set of double-wall nanotubes \cite{relief}. They found
several types of the potential relief including the type where the valleys form helical lines
analogeous to a "thread" of bolt. However, the barriers to the relative motion of walls along the
helical thread line and for jumps on neighbouring helical lines were not calculated until now.

Here, the structure of walls is constructed by folding of graphene sheet with the bond length equals
to 1.42 \AA ~(the bond length of many-wall nanotubes coincides with the bond length of graphite
within the accuracy of the neutron diffraction measurements 0.01 \AA ~\cite{bond}). The interwall
interaction is adopted here to be 6-12 Lenard-Jones potential
$U=4\epsilon((\sigma/r)^{12}-(\sigma/r)^{6})$ with parameters $\epsilon=2.968$ meV and
$\sigma=3.407$ \AA. These parameters of the potential were fitted to the interlayer distance and
modulus of elasticity of graphite and used to study the ground state and phase transition in solid
C$_{60}$ \cite{lj} and recently by Dresselhaus {\it et. al.} in study of the potential relief of
double-wall nanotubes \cite{relief}. The upper cutoff distance of the potential $10\sigma$ is used.
This cutoff distance defines the accuracy of barriers to the wall relative rotation and sliding
within 0.001 meV per atom without visible discontinuities of potential relief. The length of outer
wall $H$ corresponds to an integer number of lattice constants of this wall, namely, 1--10, 1--11,
and 1--30 lattice constants for the (6,4)@(16,4), (8,2)@(12,8), and (8,2)@(17,2) nanotubes,
respectively. The length of the inner wall is chosen so that all pairs of atoms with interatomic
distances within the cutoff distance are taken into consideration. The walls are considered to be
rigid. Account of the deformation of walls is not essential for the shape of the potential relief
both for double-wall carbon nanotubes \cite{along} and nanoparticles \cite{my}. The barriers to the
relative wall rotation and sliding for the (5,5)@(10,10) nanotube calculated here for walls with
unannealed structure coincide within 20 \% with results of Dresselhaus {\it et. al.} (used annealed
wall structure) \cite{relief}.

\begin{figure}
\includegraphics[height=8cm]{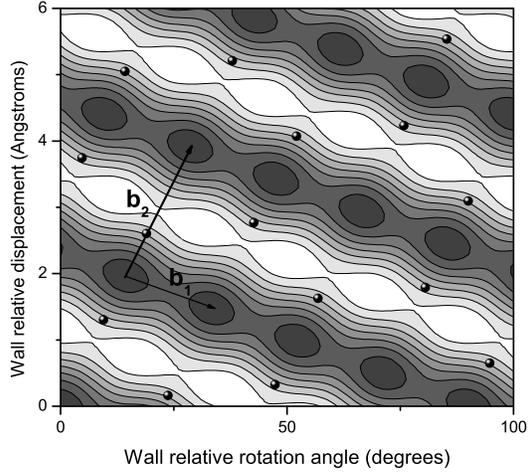}
\caption{\label{fig2_mee} The potential relief of the interwall interaction energy of the
(6,4)@(16,4) nanotube as a function of the relative displacement of the wall along the nanotube axis
and the angle of the relative rotation of the wall about the nanotube axis; ${\bf b}_1$ and ${\bf
b}_2$ are the unit vectors of the lattice formed by the minima of the potential relief. The energy
is measured from its minimum. The equipotential lines are drawn at an interval $10^{-2}$ meV per
atom. The spheres present the atoms positions of the inner wall in cylindrical coordinate system
with the axis $Z$ coincides with the axis of nanotube. The relative position of the coordinate
systems corresponding to the potential relief and atomic structure of outer wall is arbitrary.}
\end{figure}

\begin{figure}
\includegraphics[height=8cm, width=18cm]{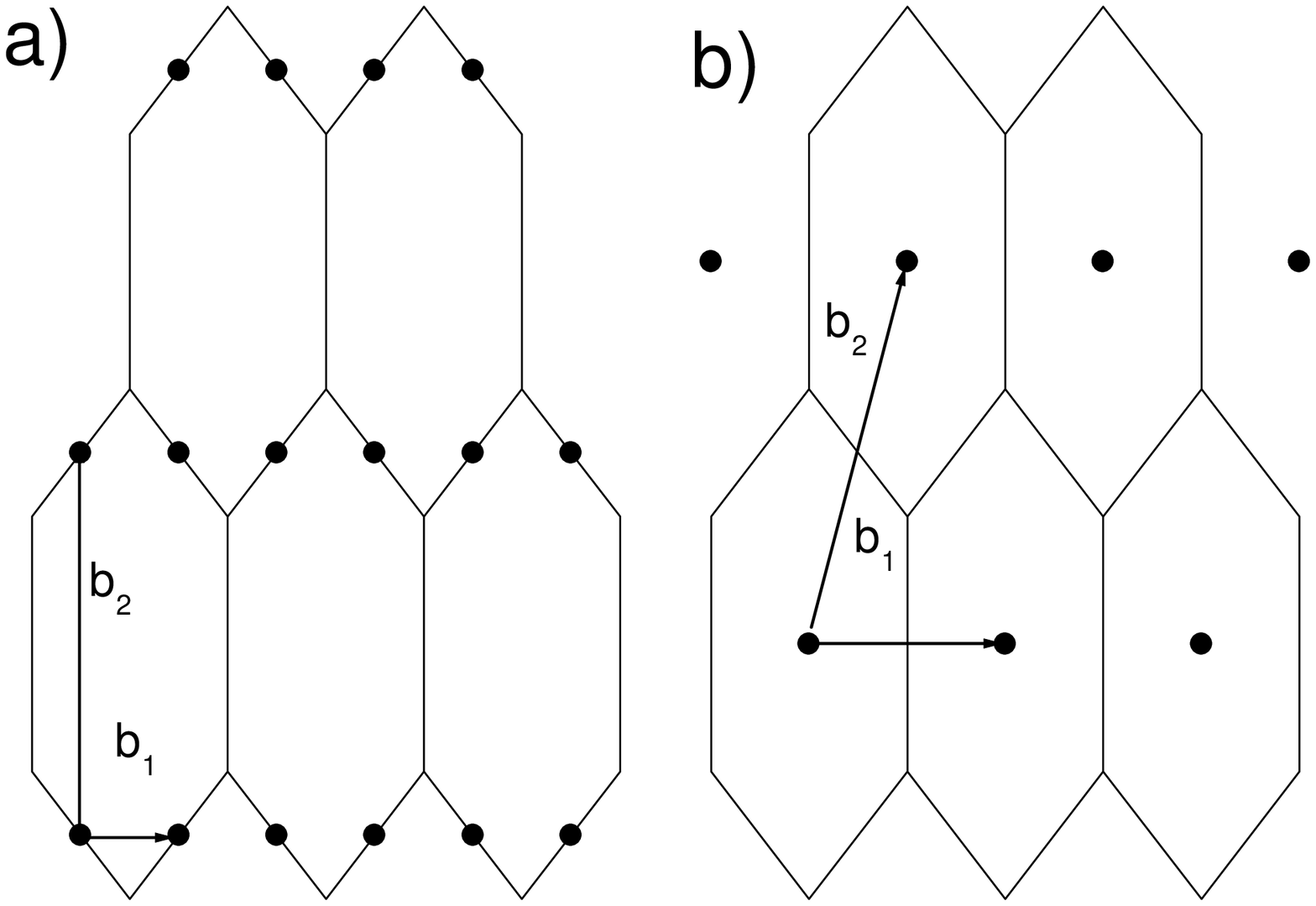}
\caption{\label{fig3_mee} The lattice of minima of the potential relief of the interwall interaction
energy (the minima positions are shown by the filled circles); ${\bf b}_1$ and ${\bf b}_2$ are the
unit vectors of the lattice formed by minima of the potential relief. Thin lines of the network
structure of the inner wall are only guides for eye that illustrate the network structure
correlation with the lattice of minima. a) type I of the potential relief, b) type II of the
potential relief.}
\end{figure}

We have studied the potential relief for the (6,4)@(16,4), (8,2)@(12,8), and (8,2)@(17,2) nanotubes,
corresponding to the case of the potential relief with the helical lines of "thread" \cite{relief}.
For the long inner wall only hexagons that are far from the wall edge take part in the walls
interaction. Since such hexagons of the network are equivalent, than the same positions of the short
outer wall relative to any hexagon of the long inner wall are also equivalent. To the contrary, the
hexagons of the short outer wall are not equivalent due to the different distances to the wall edge.
For this reason, the short outer wall can be considered as one large molecule "adsorbed" on the
surface of the inner wall. Consequently, the potential relief reproduces only the structure of the
long inner wall. Thus we have plotted the potential relief for the (6,4)@(16,4) nanotube and the
structure of the inner wall network on the same figure (see Fig. 2). Evidently that the minima of
potential relief form the lattice that correlates with the lattice of the inner wall network. Such
correlation is observed for all considered nanotubes. Two types of the potential relief minima
lattices are found. The type I, shown in Fig. 3a, corresponds to the rectangular lattice of minima.
Such a relief have the (6,4)@(16,4) (see Fig. 2) and (8,2)@(12,8) nanotubes. The type II of
potential relief, shown in Fig. 3b, corresponds to the oblique lattice of minima with the lattice
vectors having the equal lengths and angle between lattice vectors equals to 60$^o$. Such a relief
has (8,2)@(17,2) nanotube. Let us convert the angle coordinate to the length $\phi=L/R_1$, where $L$
represents the length of projection of any wall atom path along the wall surface on the wall
circumference, $R_1$ is the inner wall radius. In this coordinate system the lattice vectors have
the lengths $b_1=a_0/2$ and $b_2=\sqrt{3}a_0/2$ for the type I of the potential relief and
$b_1=b_2=a_0$ for the type II of the potential relief. For both types of the potential relief the
angle $\chi$ between the thread line and the wall circumference equals to the chirality angle
$\theta$ of the inner wall.

The barriers $U_1$ and $U_2$ between neighbouring minima in the directions along the lattice vectors
${\bf b}_1$ and ${\bf b}_2$ are different. The threadlike pattern of the potential relief arise due
to the essential difference between barriers $U_1$ for relative motion of walls along the thread
line and $U_2$ for transition of the system to neighbouring thread line ($U_1 \ll U_2$). The
nanotubes considered are incommensurate. Therefore the threadlike pattern of the potential relief
change with changing of outer wall length. The barriers to any kind of relative motion of
incommensurate walls fluctuate near their average value analogeously to the sum of functions $\cos
l$, where $l$ is an integer \cite{along}. That is the barriers for the long nanotubes are the same
order of magnitude as the analogeous barriers for the nanotubes with the outer wall length to be
equal to few unit cell lengths. The dependencies of barriers $U_1$ and $U_2$ on the outer wall
length are shown in Fig. 4 and Fig. 5, respectively. One can see that the barriers change by order
of magnitude for all nanotubes considered, while the outer wall length changes by only few
nanometers. Note that these dependencies for both barriers, at least for two of three considered
nanotubes, are quasiperiodic functions.

\begin{figure}
\includegraphics[height=8cm]{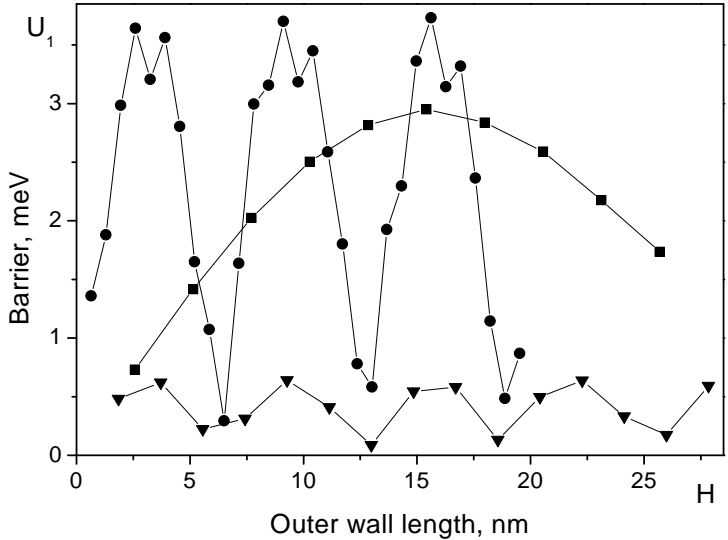}
\caption{\label{fig4_mee} The dependence of the barrier $U_1$ for relative motion of walls along the
thread line on the length $H$ of outer wall. Filled circles, filled triangles and filled squares
correspond respectively to (6,4)@(16,4), (8,2)@(12,8) and (8,2)@(17,2) nanotubes.}
\end{figure}

\begin{figure}
\includegraphics[height=8cm]{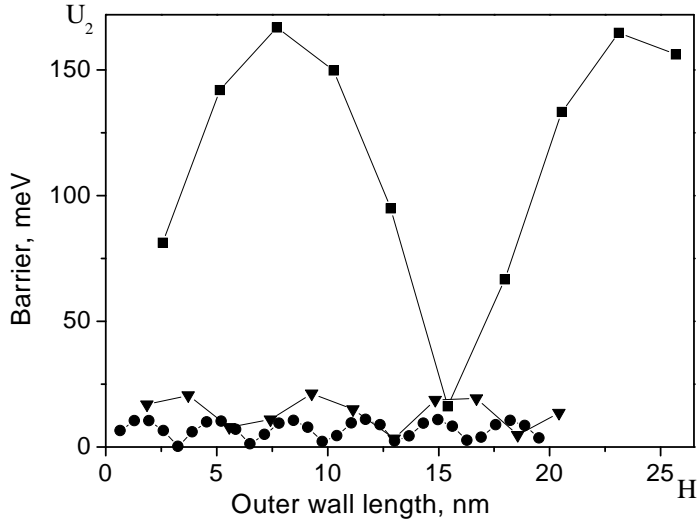}
\caption{\label{fig5_mee} The dependence of the barrier $U_2$ for transition of the system on
neighbouring thread line on the length $H$ of outer wall. Filled circles, filled triangles and
filled squares correspond respectively to (6,4)@(16,4), (8,2)@(12,8) and (8,2)@(17,2) nanotubes.}
\end{figure}

As was discussed above the quantity that characterize the possibility for double-wall nanotube to
have threadlike pattern of potential relief is not barrier itself but rather ratio $\gamma =
U_2/U_1$ of barriers to the motion across the thread line and along it, respectively. It is
naturally to call this ratio as relative thread depth. The dependence of relative thread depth on
outer wall length is shown in Fig. 6. If average periods of mentioned quasiperiodic functions are
close and oscillations of functions are in phase for both barriers then the relative thread depth
can be large for essential changes of outer wall length. The example of such a possibility is
(8,2)@(12,8) nanotube. In this case fabrication of double-wall nanotube with given relative thread
depth can be possible. If average periods of mentioned quasiperiodic functions are not close or
oscillations of function are not in phase then changes in relative thread depth under changes of
outer wall length can be of orders of magnitude. The examples of such a possibility are (6,4)@(16,4)
and (8,2)@(17,2) nanotubes. Moreover the relative thread depth for (6,4)@(16,4) nanotube is less
than 1 at some lengths of outer wall. This means disappearance of threadlike relief or possible
appearance of threadlike relief with different parameters.

***********************8

\begin{figure}
\includegraphics[height=8cm]{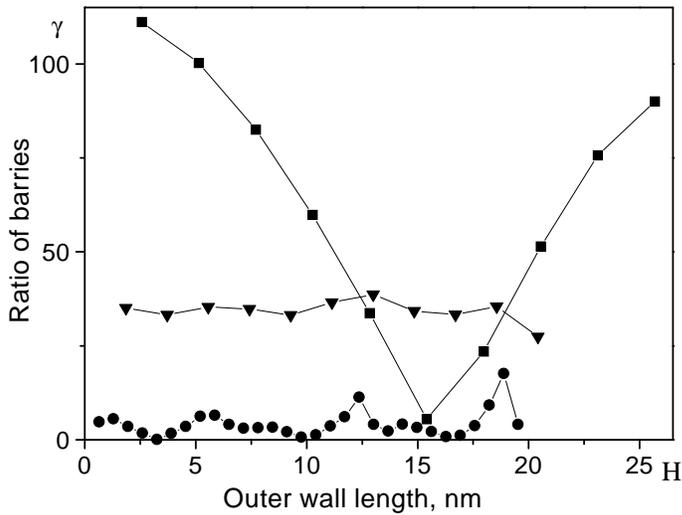}
\caption{\label{fig6_mee} The dependence of the ratio
$\gamma=U_2/U_1$ characterizing the "thread" depth on the length
$H$ of outer wall. Filled circles, filled triangles and filled
squares correspond respectively to (6,4)@(16,4), (8,2)@(12,8) and
(8,2)@(17,2) nanotubes.}
\end{figure}

Thus our calculations show that both  for relative motion of walls along the thread line and for
transition of the system to neighbouring thread line, respectively, barriers change by order of
magnitude for all nanotubes considered as the outer wall length changes by only few nanometers.
Therefore it is difficult to fabricate the double-wall nanotube with incommensurate walls but with
{\it controlled} values of barriers $U_1$ and $U_2$. Thus nanotubes with incommensurate walls are
too hard to use in nanomachines based on relative motion of walls where the {\it precise control} of
relative position of walls is necessary. However such nanotubes can be used in nanomachines based on
fast relative motion of walls where only the presence of "threadlike" potential relief with high
value of relative thread depth $\gamma$ is necessary.

The barriers to any kind of relative motion of commensurate walls with lengths corresponding to
integer number of nanotube elementary cells are given by relation $U_a=U_u N_u$, where $U_u$ is the
barrier per unit cell of the nanotube, $N_u$ is a number of unit cells in the nanotube (the
interaction with atoms on the edge of wall is disregarded here). Thus the barrier $U_a$ for
sufficiently long nanotube is proportional to its length and  can be possible to obtain the given
value of barrier by choice of nanotube length. The analysis show that for some two-wall nanotubes
with integer number of elementary  cells the barriers can be extremely small in comparison with
total interwall interaction energy. The reason the fact is following. The potential field produced
by each wall can be expanded in the basis of harmonics invariant under symmetry group of the wall
\cite{yug1}. Only harmonics with symmetry compatible with both walls can contribute in the interwall
interaction potential relief $U(\phi,z)$. As was shown on example of nonchiral wall (12,12) the
amplitudes of these harmonics sharply decrease and rapidly becomes negligible. For example, in
(7,7)@(12,12) nanotube only 24-th harmonic of expansion of potential field produced by wall (7,7)
can contribute in the angle dependence of the interwall interaction energy. Therefore it was found
that barrier for relative wall rotation for this nanotube is less than calculation accuracy
\cite{yug2}. The analogous results one can expect for the majority of nanotubes with chiral
commensurate walls.

To systemize the search of nanotubes with commensurate walls and threadlike potential relief the
notion of equivalence class of walls can be introduced (see details in \cite{bel}). An equivalence
class is the set of all walls with chirality indexes $(kf,kg)$, where $k$ is integer, $f$ and $g$
are coprime natural numbers, $k$ is called diameter index and pair $(f,g)$ is called chirality
indexes of the class. If the walls from different equivalence classes $(f_1,g_1)$ and $(f_2,g_2)$
are commensurate then the interwall distance for the nanotube $(k_1f_1, k_1g_1)@(k_2f_2, k_2g_2)$
composed of these walls is given by the formula

\begin{equation}
\Delta R=\frac{a_0}{2\pi}\sqrt{f_2^2+f_2g_2+g_2^2}(k_2 -
k_1\frac{d_1}{d_2}),
\end{equation}

\noindent Such a pair of equivalence classes generates double-wall nanotubes families with the
geometrical parameters that are the same for all nanotubes of each family: the interwall distance,
unit cell length of nanotube and difference of chirality angles of walls.

\section{Dynamics of relative motion of walls.}

Let us study now the nanomechanical problem, dynamics of relative motion
of double-wall nanotube
interacting walls under the action of external forces. The wall interaction
potential is denoted as $U(\phi,z)$, where $\phi$ is the angle of wall relative
rotation about nanotube axis and $z$ is the wall relative displacement along
nanotube axis. One wall is treated as fixed and the motion of second
wall relative to the one is examined. We consider the case where
external force causes the relative motion of walls but does not cause their
deformation. In this case forces ${\bf F}^a$ that acts on each atom
of movable wall have equal magnitudes. This forces can be divided into two components
${\bf F}^a={\bf F}^a_z+{\bf F}^a_L$ where components ${\bf F}^a_z$ and
${\bf F}^a_L$ are directed along the nanotube axis (first type forces)
and along the tangent to nanotube circumference (second type forces),
respectively. The forces of considered type can have, for
example, electrostatic nature \cite{switch} or can be applied by the nanomanipulator
\cite{tele} or by laser electric fields \cite{bearings,gear}.

The analysis shows that for case considered resultant force
${\bf F}_r$ corresponding to components ${\bf F}^a_L$ can be majorize as

\begin{equation}
   \left| {\bf F}_r \right| \le \frac{F^a_L N_c}{\pi p},
\end{equation}

\noindent where $N_c$ is number of graphite unit cell in a unit cell of
one-dimensional crystal corresponding to the movable wall, $p$ is the number
of complete revolutions of helical line determined by the wall symmetry
\cite{symmetry,class1} per unit cell of the wall. Thus for sufficiently long
wall the contribution of tangential components ${\bf F}^a_L$ in resultant force
acting on movable wall center of mass can be
disregarded. Therefore tangential components ${\bf F}^a_L$ cause only rotation
of wall and wall center of mass moves only along nanotube axis.
On using once again the transformation $\phi=L/R_1$
it is easily to show that equations of motion of center of mass
and rotation for movable wall are equivalent to one vector equation

\begin{equation}
\label{mov}
          M \ddot {\bf r} = -\frac{d U({\bf r})}{d{\bf r}} + {\bf F}
\end{equation}

\noindent where $M$ is the movable wall mass, ${\bf r}$ is vector with
components $z$ and $L$ and ${\bf F}=(N_aF^a_z,N_aF^a_L)$, $N_a$ is
number of atoms in the movable wall.

Thus the motion of one wall of double-wall carbon nanotube
relative to fixed wall in the absence of external forces or under the action of
external forces satisfying the conditions described above is equivalent to
two-dimensional motion of particle with mass $M$ in the potential field
$U({\bf r})$ and under the action of external force ${\bf F}$.

Now we shall obtain Fokker-Planck equation describing relative diffusive motion of carbon nanotube
walls and their drift under the action of force ${\bf F}$. We consider the case corresponding to the
simulations performed. Firstly let the length of movable wall is considerably less than length of
fixed wall and movable wall is placed far from fixed wall edges so that interaction of movable wall
with atoms at the edge of fixed wall has negligible effect on the potential relief. Thus we leave
aside the case where relative motion of walls is telescopic with the extension of outer wall outside
inner one. In last case contact area of walls and, consequently, corresponding surface energy
substantially depend on $z$, and, consequently, the minima of potential relief can disappear.
Secondly, let the minima $U_{min}$ of potential relief $U({\bf r})$ form the lattice with lattice
vectors ${\bf b}_1$ and ${\bf b}_2$ as it is shown in  Fig. 2. The relative motion of walls can be
diffusion with drift only in the case $k T \ll U_1,U_2$, where $U_1$ and $U_2$ are the barriers
between minima $U_{min}$ for motion along lattice vectors ${\bf b}_1$ and ${\bf b}_2$, respectively.
We restrict ourselves by the case of $U_1 \ll U_2$, that is the case of one-dimensional diffusion in
direction along lattice vector ${\bf b}_1$. In this case, if the direction of ${\bf b}_1$ coincides
with the direction of vertical z-axis or horizontal $\phi$-axis, the considered relative motion of
walls is relative sliding or rotation, respectively. Otherwise this motion is analogeous to the
relative motion of bolt-and-nut pair.

Let us consider the ensemble of "particles"
with the motion described by equation (\ref{mov}), where potential $U({\bf r})$
and force ${\bf F}$ have all properties described above. The number of particles
$dN$ passing during the time $d t$ through normal to diffusion direction
segment $d \xi$  is given by

\begin{equation}
\label{dn}
                d N = dN_1-dN_2 = \frac{1}{2} (n_1 u_1 - n_2 u_2) d \xi d t,
\end{equation}

\noindent where $dN_1$ and $dN_2$ are numbers of particles passing in opposite directions ${\bf
b}_1$ and $-{\bf b}_1$ between neighbouring minima $U_{min}$ separated by segment $d \xi$; $n_1$ and
$n_2$ are concentrations of particles at some distances $\lambda$ from segment $d \xi$ corresponding
to particles moving in directions ${\bf b}_1$ and $-{\bf b}_1$; $\lambda$ is the collision mean free
path of particles before passing through segment $d \xi$; $u_1$ and $u_2$ are average velocities of
particle displacement in directions ${\bf b}_1$ and $-{\bf b}_1$, respectively. These velocities are
given by

\begin{equation}
\label{u}
                 u_1 =\frac{\delta}{\tau_1},
                 ~~~~~~~ u_2 =\frac{\delta}{\tau_2}
\end{equation}

\noindent where $\delta$ is the distance between neighbouring minima in the motion direction,
$\tau_1$ and $\tau_2$ are average times of displacement between two neighbouring minima for motion
in directions ${\bf b}_1$ and $-{\bf b}_1$, respectively, $\tau_1=1/w_1$ and $\tau_2=1/w_2$, $w_1$
and $w_2$ are probabilities of corresponding displacements. The quantities $w_1$ and $w_2$ are not
equal in result of force $\bf {F}$ action. Namely, the barriers $U_1$ between neighbouring minima
change by the work $W=F_x \delta /2$, where $F_x$ is the projection of $\bf {F}$ on particle motion
direction. This change is decrease if $F_x$ is directed along the particle motion and increase
otherwise. The probabilities $w_1$ and $w_2$ are given approximately by Arrhenius formula

\begin{equation}
\label{w}
                 w_1=\Omega \exp \left( - \frac{U_1-F_x \delta /2}{k T} \right),
~~~~~~~~~~~~~~~~~w_2=\Omega \exp \left( - \frac{U_1+F_x \delta /2}{k T} \right),
\end{equation}

\noindent  where $\Omega$ is a frequency which has the same
order of magnitude as oscillation frequency of the particle near the
minimum.

The particle motion is diffusion if $kT \ll U_1,U_2$ and drift (that is the
time-average acceleration is zero) if $F_x \delta /2 \ll U_1$.
To produce the Fokker-Planck equation the first term of exponent
$\exp(F_x \delta/ 2 k T)$ expansion must be used, therefore the condition
$F_x \delta /2 \ll k T$ is also necessary.

Substituting (\ref{u})--(\ref{w}) in (\ref{dn}), we get

\begin{equation}
\label{dn2}
        d N = \frac{1}{2} \Omega \delta \exp \left( - \frac{U_1}{k T} \right)
\left[ (n_1 - n_2)+
 (n_1 + n_2)\frac{F_x \delta /2}{k T} \right] d \xi d t
\end{equation}

With the substitutions $n_1 + n_2 \approx n$ and
$n_1 - n_2 \approx \delta \partial n /\partial x$ equation (\ref{dn2}) takes the form of
Fokker-Planck equation:

\begin{equation}
\label{dn3} \frac{\partial n}{\partial t} = D \frac{\partial^2 n}{\partial x^2} + \frac{\partial
n}{\partial x} B F_x
\end{equation}

Here $D$ and $B$ are, respectively, diffusion coefficient and
mobility of particles given by

\begin{equation}
                D =  \frac{1}{2} \Omega \delta^2 \exp \left( - \frac{U_1}{k T} \right)
\end{equation}

\begin{equation}
                B =  \frac{\Omega \delta^2}{2k T} \exp \left( - \frac{U_1}{k T} \right)
\end{equation}

\noindent Note, that Einstein ratio $D=k T B$ is fulfilled.

The behaviour of one particle with fixed coordinates at $t=0$
is described by solution $n_s(x,t)$ of the equation (\ref{dn3}) that satisfy two
conditions: 1) at $t=0$ all the concentration is placed on a line
normal to the diffusion direction and the concentration elsewhere on the
plane is zero; 2) the total number of particles is equal to 1.

\section{Discussion}

We consider here two types of nanomachines based on relative motion of nanotube walls. Let us
discuss firstly possible advantages of application of nanotubes with threadlike potential relief of
the interwall interaction energy in nanomachines where the direction of forces applied on movable
wall does not correspond to desirable kind of wall motion. In the case where the potential relief
has negligible effect on relative motion of walls, the directions of forces applied on movable wall
must correspond to direction of relative motion of walls. Namely, if the relative motion of walls is
sliding along nanotube axis, as it take place in constant-force nanosprings \cite{tele}, gigahertz
oscillators \cite{giga}, and mechanical nanoswitch \cite{switch}, than the forces applied on movable
wall are bound to be directed along nanotube axis (first type forces). If the relative motion of
walls is relative rotation, as it take place in nanobearings \cite{bearings} and nanogears
\cite{gear} than the forces applied on movable wall must to be directed along the tangent to its
circumference (second type forces).

However the presence of threadlike potential relief of the interwall interaction energy remove the
restriction on directions of forces applied on movable wall. The analysis above shows that relative
motion of walls along helical line of "thread" is possible for both discussed types of external
forces and any their superposition. Therefore the first type forces produce not only a relative
sliding of walls along axis but also their relative rotation. Therefore a nanomachine based on wall
motion can operate as a {\em nanowhirligig}. The proposed way to convert the forces directed along
nanotube axis into relative rotation of walls can be used in nanobearings and nanogears. The second
type forces producing a rotational moment gives rise not only a relative rotation but also a
relative motion of walls along nanotube axis. This effect provides a possibility to construct a
nanomachine based on carbon nanotube that is analogeous to old-desinged {\em faucet} where rotation
of handle converts into forward motion of rod.

We propose here also principally new type of nanomachines that may be based on nanotubes with only
threadlike potential relief of the interwall interaction energy. The using of alternating-sign force
to operate the relative position of walls can produce a motion of walls that is analogeous to the
motion of auger in a perforating drill. Such a {\em perforating nanodrill} can be used for
modification of surface in nanometer size (see Fig. 7B).

\begin{figure}
\includegraphics[width=15cm]{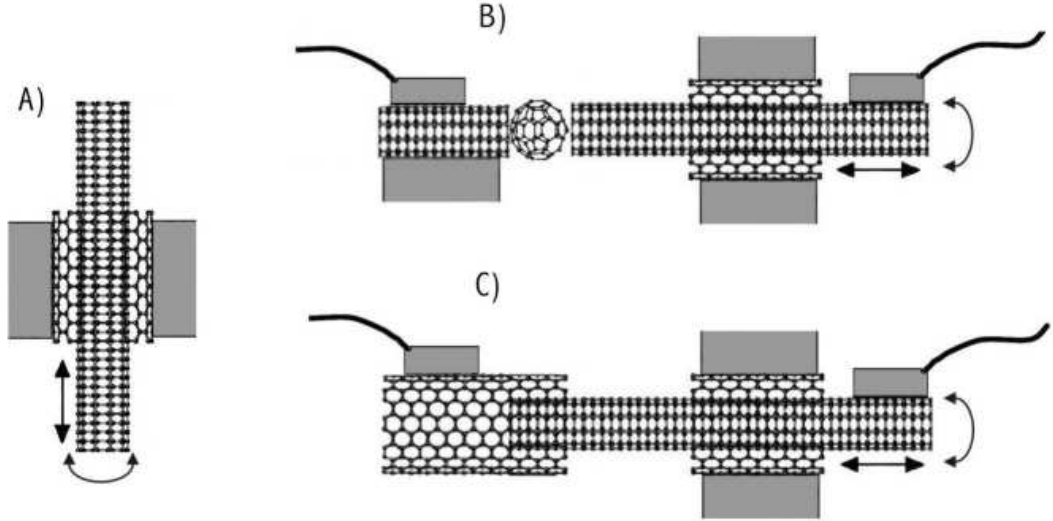}
\caption{\label{fig7} Nanomachines based on a threadlike relative motion of nanotube
walls: A) nanodrill, B) and C) two types of variable nanoresistors.}
\end{figure}

Another new type of nanodevices which we propose here are based on relative motion of nanotube walls
are electromechanical nanodevices. Two nanometer size systems with the conductivity which can be
tuned within orders of magnitude by the rotation of a nanotube wall or its displacement along the
axis were considered. The first one consists of two single-wall nanotubes and a fullerene between
them \cite{switch1} (see Fig. 7B), and the second one is an telescoping nanotube \cite{telero} (see
Fig. 7C). The tuning can be controlled with the help of nanodevice based on relative motion of
nanotube walls. As a result {\it a variable nanoresistor} can be constructed, where nanotube wall is
both movable element and element of the electric circuit.

Another new type of nanodevices which we propose here based on relative motion of nanotube walls are
electromechanical nanodevices. For example, the conductivity of system consisting of two carbon
nanotubes and a fullerene between them \cite{switch1} can be tuned within orders of magnitude by
rotation of a one nanotube or its displacement along the axis. This tuning can be controlled with
the help of nanodevice based on relative motion of nanotube walls. In result a {\em variable
nanoresistor} can be constructed, where wall of nanotube is both movable element and element of
electric circuit.

Let us discuss possible operation modes of nanomachines based on relative motion of nanotube walls
along line of thread. As we have shown above, in the case, when conditions $k T \ll U_1,U_2$ and
$F_x \delta /2 \ll k T$ are fulfilled, the relative motion of carbon nanotube walls is described by
Fokker-Planck equation (\ref{dn3}). The operation mode of nanomachine based on such a motion is
called here Fokker-Planck operation mode. This mode is worthwhile to use in nanomachine if average
distance $x_{dr}=B F_x t$ passed by a wall along a helical line of "thread" in result of drift is
greater than average distance $x_{dif}=\sqrt{2Dt}$ passed by this wall in result of diffusion. This
condition is fulfilled for displacements $x_{dr} \gg \delta$, that is, e.g., for tens of relative
jumps of wall along a helical thread line between minima of the interwall potential $U({\bf r})$.
Such displacement along a helical line corresponds to less than one revolution of wall about the
nanotube axis or displacement by nanometeres along this axis. Although Fokker-Planck operation mode
does not allow the precise control of relative positions of walls, this mode can be used, for
example, in perforating nanodrill for perforation of layers with thickness less than average
displacement $x_{dr}$ of wall (that plays the role of auger) in result of drift.

For forces $F_x \delta /2 \gg k T$ the stochastic contribution in relative motion of
walls can be neglected. In this case the relative motion of walls is accelerated and as
it is discussed above equivalent to two-dimensional motion of one particle
described by Eq. (\ref{mov}). The
operation mode of nanomachine based on such a motion is called here the accelerating
operation mode. In this mode the controlled relative displacement of walls along a
helical line of "thread" for distance that is less than $\delta$ is possible. This mode
can be used, for example, in variable nanoresistor.

For {\it intermediate} forces $F_x \delta /2 \approx k T$ the next terms of exponent
expansion can be used. Then Fokker-Planck equation (\ref{dn3}) takes the analogeous
form as for small forces

\begin{equation}
\frac{\partial n}{\partial t} = D' \frac{\partial^2 n}{\partial x^2} + \frac{\partial
n}{\partial x} B' F_x
\end{equation}

\noindent where the diffusion coefficient $D'$ and the mobility $B'$ are expressed in
terms of diffusion coefficient $D$ and mobility $B$, in the case of small forces,
respectively.

\begin{equation}
                D' =D  \left( 1 + \frac{F^2_x \delta^2}{8 k^2 T^2} \right)
\end{equation}

\begin{equation}
                B' =B  \left( 1 + \frac{F^2_x \delta^2}{24 k^2 T^2} \right)
\end{equation}

Let us estimate the range of forces that can be used to control the relative motion of
carbon nanotube walls in nanomachine operating in Fokker-Planck and in accelerating
operation modes. Our estimations are made for nanotube (8,2)@(12,8). The ratio of
barriers $\gamma=U_1/U_2$ of this nanotube conserves within the range 25-40 for all
considered lengths of outer wall. The conditions $k T \ll U_1,U_2$ and $F_x \delta /2
\ll k T$ give the maximal force $F_{FP}$ corresponding to Fokker-Planck mode $F_{FP}
\ll 2U_1/\delta$, where

\begin{equation}
\delta = \frac{a_0}{2} \sqrt{\left( \frac{R_2}{R_1} \right)^2 \cos^2 \chi + \sin^2
\chi},
\end{equation}

\noindent Here $R_1$ and $R_2$ are radii of inner and outer wall, respectively, $\chi$
is the angle between helical line of "thread" and wall circumference.

For nanotube (8,2)@(12,8) we have: 1) value of $\chi$ equal to that of chiral angle
$\theta=10.89^o$ and 2) magnitude of $U_1 \approx 0.6$ meV corresponding to outer wall
length that equals to the length of unit cell of the wall. In result we get $\delta =
1.86$ \AA ~and $F_{FP} \ll 10^{-12}$ N.

The using of too high force controlling relative motion of walls at accelerating mode can give rise
twist-off. The twist-off can occur only if the projection $F_y$ of external controlling force on the
direction normal to thread line will satisfy to the inequality

\begin{equation}
\label{fa} F_y > \left\langle \frac{\partial U(y)}{\partial y} \right\rangle _y \approx
\frac{2U_2}{\delta_y}
\end{equation}

\noindent where $y$ is wall relative displacement in the direction normal to thread line and
$\delta_y$ is the distance between neighbouring lines of "thread"

\begin{equation}
\delta_y = \frac{\sqrt{3} a_0}{2} \sqrt{\left( \frac{R_2}{R_1} \right)^2 \cos^2{\chi} +
\sin^2 {\chi}}
\end{equation}

\noindent For controlling forces greater than $F_{ac} = 2U_2/\delta_y$ the relative motion of walls
in the direction normal to thread line must be taken into account. For controlling forces less than
$F_{ac}$ it is sufficiently to consider the relative motion of walls only along the thread line. For
nanotube (8,2)@(12,8) on substituting in Eq. (\ref{fa}) $\delta_y = 2.23$ \AA, $U_2= 20$ meV and
magnitude of $\chi$ equal to chiral angle $\theta$ we get $F_{ac} \approx 3 \cdot 10^{-11}$ N.

\section*{Acknowledgements}

Yu.E.L. is grateful to S.V. Iordanskii and A.Ya. Vul for useful discussions. This work
was supported by grants of Russian Foundation of
Basic Researches and Ministry of Sciences.\\

\end{document}